\documentclass[prb,aps,amsmath,showpacs,preprint,]{revtex4-1}
\usepackage{graphicx}
\usepackage{epsfig}
\usepackage{multirow}
\usepackage{dcolumn}

\usepackage{bm}
\usepackage{subfigure}
\usepackage{color}

\begin{document}

\title{Thermoelectric properties of marcasite and pyrite FeX$_2$(X=Se,Te): A first principle study}

\author{Vijay Kumar Gudelli and V. Kanchana$^{*}$}
\affiliation {Department of Physics, Indian Institute of Technology Hyderabad, Ordnance Factory Estate, Yeddumailaram-502 205, Andhra Pradesh, India}
\author {G. Vaitheeswaran}
\affiliation {Advanced Centre of Research in High Energy Materials (ACRHEM), University of Hyderabad, Prof. C. R. Rao Road, Gachibowli, Hyderabad-500 046, Andhra Pradesh, India}
\author{M. C. Valsakumar}
\affiliation{School of Engineering Sciences and Technology (SEST), University of Hyderabad, Prof. C. R. Rao Road, Gachibowli, Hyderabad-500 046, Andhra Pradesh, India}\author {S. D. Mahanti}
\affiliation {Department of Physics and Astronomy, Michigan State University, East Lansing, Michigan 48824, USA}
\date{\today}

\begin{abstract}
Electronic structure and thermoelectric properties of marcasite (m) and synthetic pyrite (p) phases of FeX$_2$ (X=Se,Te) have been investigated using first principles density functional theory and Boltzmann transport equation. The plane wave pseudopotential approximation was used to study the structural properties and full-potential linear augmented plane wave method was used to obtain the electronic structure and thermoelectric properties (thermopower and power factor scaled by relaxation time). From total energy calculations we find that m-FeSe$_2$ and m-FeTe$_2$ are stable at ambient conditions and no structural transition from marcasite to pyrite is seen under the application of hydrostatic pressure. The calculated ground state structural properties agree quite well with available experiments. From the calculated thermoelectric properties, we find that both m and p forms are good candidates for thermoelectric applications. However,  hole doped  m-FeSe$_2$ appears to be the best among all the four systems.
\end{abstract}

\maketitle

\section{Introduction}
The performance of a thermoelectric (TE) material depends on the dimensionless figure of merit, ZT, given by $\frac{S^2\sigma T}{\kappa}$, where S, $\sigma$, T and $\kappa$ are the Seebeck coefficient, electrical conductivity, absolute temperature and the thermal conductivity (which includes both electronic $\kappa_e$ and lattice contribution $\kappa_l$, i.e. $\kappa=\kappa_e+\kappa_l$) respectively, the efficiency of a thermoelectric device increasing with ZT. The best of the commonly available TE materials have a value of ZT to be of the order of unity.\cite{Bi2Te3} From the above expression for ZT, it is evident that finding materials with high ZT (more than unity) still remains an open challenge, as it needs to satisfy the conflicting requirement of high thermopower like an insulator and behave as a good conductor like a metal. Also it implies the need for materials with good electrical conductivity and poor thermal conductivity resulting in weak electron scattering and strong phonon scattering. In last few years efforts have been made for identifying strategies to improve the value of the ZT. Several reports have been published by different research groups with focus on band structure engineering to enhance S and $\sigma$ and usage of nanostructure technology for reducing the lattice thermal conductivity. Recently, Xun Shi et al,\cite{Shi} reported that the multiple-filled skutterudites show an improved figure of merit of 1.7 at 850 K, which is the highest value reported in skutterudites. Biswas et al.,\cite{Biswas} reported that PbTe-SrTe doped with Na shows a ZT of 2.2 at 923 K due to the hierarchical structure which maximises the phonon scattering.

\par There are well known constraints in developing good TE materials, like toxicity and scarcity of the elements which prevent the usage of above materials in large scale industrial application. Nevertheless, the search for such new TE materials still continues despite the above mentioned restrictions.
Recently, the natural minerals of the tetrahedrite (Cu$_{12-x}$M$_x$Sb$_4$S$_{13}$) and tennantite (Cu$_{12-x}$M$_x$As$_4$S$_{13}$), where M is a transition metal element such as Zn, Fe, Mn, or Ni, have shown potential thermoelectric application due to their intrinsic low lattice thermal conductivity.\cite{Slack,Xlu_1,Xlu_2} Such studies have motivated us to explore thermoelectric properties of other family of minerals such as FeSe$_2$ and FeTe$_2$. The reason behind selecting the transition metal chalcogenides family is due to their excellent optical and magnetic properties,\cite{CNR} and the potential for widespread applications. Recently, the polymorphic phases of FeSe$_2$ have been shown to be good for solar cell absorber application.\cite{Ganga} Several experimental reports are available attempting to understand the electrical resistivity, Hall coefficient and thermoelectric power of these compounds. The electrical resistivity and Hall coefficient of FeSe$_2$ have been measured in sintered poly-crystals.\cite{Fischer,Dudkin,Hulliger} Dudkin {\it et. al.,} have measured the electronic resistivity of FeTe$_2$. The same authors have also reported the thermoelectric properties of FeSe$_2$ and FeTe$_2$, measured at ambient temperature and the high temperature results have been reported by Harada.\cite{Harada} The thermoelectric properties of pyrite-type FeSe$_2$ and FeTe$_2$ prepared at high pressure of 65 kbar was given by Bither {\it et. al.}\cite{TAB} There are no theoretical studies on these compounds to understand  thermoelectric properties.

\par In this work, we present a detailed theoretical study of electronic structure and thermoelectric properties of both the marcasite and pyrite phases of FeX$_2$(X=Se,Te), for which the available experimental data are indicative of good TE potential, which however is not realized so far. The paper is organized as follows: in section II we describe the method used for the calculations and section III presents the results and discussion, followed by conclusion in section IV.

\section{Methodology}
All the total energy calculations based on first principle density functional theory (DFT) were performed using pseudopotential method as implemented in the Plane wave self-consistent field (Pwscf) program \cite{Pwscf} and full-potential linear augmented plane wave (FP-LAPW) method as implemented in the WIEN2k.\cite{Blaha} The Pwscf method is used to perform the structural optimization, whereas FP-LAPW method is used to study the electronic and transport properties. The total energies are obtained by solving the Kohn-Sham equations self consistently within the Generalized Gradient Approximation (GGA) of Perdew-Burke-Ernzerhof (PBE).\cite{Perdew} A plane wave kinetic energy cut off of 50 Ry is used and the first Brillouin zone in the reciprocal space is sampled according to the Monkhorst-Pack scheme \cite{Monkhorst} by means of a 8x8x8 k-mesh in order to ensure that the calculations are  well converged. The electron-ion interactions are described by Vanderbilt type ultrasoft pseudo potentials \cite{Vanderbilt} and the following basis sets Fe: 3s$^2$ 3p$^6$ 3d$^6$ 4s$^2$, Se: 4s$^2$ 4p$^4$ and Te: 5s$^2$ 5p$^4$ are used as valence states. Variable-cell structural optimisation has been performed by using BFGS (Broyden-Fletcher-Goldfarb-Shanno) conjugate gradient algorithm as implemented in Pwscf. To determine the ground state structure of FeSe$_2$ and FeTe$_2$ and possible phase transformation, we have calculated the total energy with applied hydrostatic pressure for both marcasite and pyrite crystal structures ranging from -8 GPa (expansion) to 8 GPa (compression) with a step size of 1 GPa.
The threshold criteria of 1 $\times 10^{-5}$ Ry for total energies, 1 $\times 10^{-4}$ Ry/bohr for total forces and 0.002 GPa for total stress were used for structural relaxation at each pressure.

\par To study the electronic properties, we have used FP-LAPW method as implemented in the WIEN2k code. As it is well known, the first principles calculations often underestimate the band gap within the standard local scheme of the exchange-correlation functional (LDA or GGA) and they also fail to describe accurately the localised electrons in {\it d} or {\it f} states, in transition metal and rare earth compounds.\cite{Nieminen} In order to overcome these drawbacks of the standard exchange methods, we have used the Generalized Gradient Approximation (GGA) along with the onsite Coulomb repulsion U (GGA+U). Here we have used 1000 k-points for calculating the electronic structures of both marcasite and pyrite forms. All our calculations are performed using the optimized parameters from the Pwscf calculation with an energy convergence up to 10$^{-6}$ Ry  per unit cell between the successive iterations. Further we have calculated the properties like thermopower (S), electrical conductivity ($\frac{\sigma}{\tau}$) using BOLTZTRAP\cite{Madsen} code with as many as 100 000 k-points, within the Rigid Band Approximation (RBA)\cite{Scheidemantel,Jodin,Chaput} and the constant scattering time ($\tau$) approximation (CSTA). According to the RBA approximation, doping a system does not alter its band structure but varies only the chemical potential, and it is a good approximation for doped semiconductors to calculate the transport properties theoretically when doping level is not very high.\cite{Jodin,Chaput,Bilc,Ziman,Nag} However certain types of dopant can drastically modify the nature of electronic structure near the gap giving rise to resonant states\cite{Bilc_04,Ahmad} in which case the RBA can fail.\cite{MSlee} According to CSTA, the scattering time of the electron is taken to be independent of energy and depends only on concentration and temperature. The detailed explanation about the CSTA is given in Ref. \onlinecite{singh, aggate2,  Khuong} and the references cited therein. It is evident that CSTA has been quite successful in the past in predicting the thermoelectric properties of many materials.\cite{DJS, Parker, Georg, Khuong, Lijun}

\section { Results and discussion }

\subsection{Ground state properties}
FeSe$_2$ and FeTe$_2$ crystallize in both the marcasite and the pyrite structures.\cite{Bither} The marcasite form of both the compounds are available in nature whereas pyrite structure is a synthetic mineral. The atomic arrangements of the marcasite phase can be considered within either of the two space groups {\it Pnn2} or {\it Pnnm}. However, we did not find any significant energy difference between these two arrangements (see Fig. 1(b) for FeTe$_2$).\cite{Gunnar} In general, most marcasite type minerals are available in the space group {\it Pnnm}, and hence we have used this space group for detailed electronic structure calculations for both the compounds. In order to verify their structural relation between the marcasite and pyrite we have calculated the total energy under the application of the hydrostatic pressure from -8 to 8 GPa. The total energy variation with the pressure for both the compounds is shown in Fig. 1(a) and (b). We find an energy difference of 1.9 mRy/unit cell between the marcasite and pyrite structures of FeSe$_2$, whereas we found the energy difference between the marcasite and pyrite phases of FeTe$_2$ to be 3.5 mRy/unitcell (see Fig. 1b).
The optimized structural parameters are shown in Table-I along with available experimental results. The agreement between theory and experiment is quite good.

\subsection{Band structure and Density of states}
The electronic properties of FeX$_2$ (X=Se,Te) are calculated using the optimized parameters obtained from the Pwscf calculations. Since LDA/GGA underestimate the band gaps in semiconductors and insulators, and as the studied compounds have partially filled Fe d-levels, we used GGA+U method and adjusted U to get a reasonable value of the band gap. In our calculations we have used  a value of U$_{Fe}$ = 0.52 Ry (7.07 eV) for the Fe {\it d} electrons in both the structures and both the compounds. This value of U is slightly on the higher side compared to values used in the literature (3-5 eV) for metals and semiconductors. The calculated band structures along high symmetry directions in the Brillouin zone for both the compounds and both marcasite and pyrite structures are shown in Figs. 2-5, along with the density of states (DOS).
\par Both the compounds are indirect band gap semiconductors irrespective of their crystal structures. From the partial density of state (PDOS) analysis, we find that there is a strong hybridization between Fe-{\it d} and chalcogen {\it p}-bands. The Fe {\it d}-bands are partially filled and Se {\it p}-bands are partially empty. The top of the valence band is predominantly Fe-{\it d} whereas the bottom of the conduction band is predominately chalcogen {\it p}. However a closer look at the PDOS shows that the states within an energy range 0.25 eV just below the valence band maxima (VBM) (responsible for charge and energy transport) are equal mixture of Fe-{\it d} and chalcogen {\it p} states. In contrast, the states near the bottom of the conduction band minima (CBM) are mostly of chalcogen {\it p}-character.  In the marcasite phase of FeSe$_2$ (m-FeS2), the CBM and VBM are located along the $\Gamma$ - Y and $\Gamma$ - X directions respectively whereas for FeTe$_2$ (m-FeTe$_2$) they are along the  $\Gamma$ - Y and $\Gamma$ - X directions, although there is another CBM along the $\Gamma$ - Z direction nearby in energy. In contrast, in the pyrite phase both of them (p-FeSe$_2$ and p-FeTe$_2$) show similar behavior, CBM is at the $\Gamma$ point and the VBM lies along the $\Gamma$ - M direction. Quite interestingly, in p-FeTe$_2$, there are other nearly degenerate VBM along the $\Gamma$ - X directions. The nature of CBM and VBM and states near their neighborhood will have significant effect on the thermoelectric properties of these two compounds, as discussed later in the paper.

\par The theoretical values of the band gaps are 1.23 eV for m-FeSe$_2$ and 0.69 eV for p-FeSe$_2$, in good agreement with earlier theoretical calculations by Ganga {\it et al}\cite{Ganga} given in Table-II. The corresponding band gaps in the Te compounds are respectively 0.33 and 0.34 eV. The overall reduction in band gap in tellurides is consistent with other known chalcogenides (Bi$_2$Se$_3$, Bi$_2$Te$_3$ etc where the gap decreases in going from Se to Te). However, the sensitivity of the band gap to the structure in FeSe$_2$ and lack thereof in FeTe$_2$ is an important difference between these two compounds. As regards comparison with experiment (see TABLE II), theoretical values of the band gap in m-FeSe$_2$ (1.234 eV in this work using GGA+U and 0.86 eV by Ganga et al using GGA) are in reasonable agreement with experiment (0.95-1.03 eV). GGA underestimates whereas GGA+U overestimates the band gap. However in m-FeTe$_2$ the discrepancy between theory (0.328 eV using GGA+U) and experiment (0.92 eV) is quite large and in the wrong direction compared to m-FeSe$_2$.  We expect experimental band gap in m-FeTe$_2$ should be smaller than that of m-FeSe$_2$. In view of this we are calling for more experiments on the optical properties on FeTe$_2$ to measure its band gap. To further understand the difference between the two compounds we have calculated the effective masses in the neighborhood of different VBM and CBM. The calculated results are shown in Table-III. Rapid increase in the DOS near the CBM in the marcasite phase for both the compounds suggests that these will be excellent n-type thermoelectric. In contrast, the pyrite structure is more favorable to p-type thermoelectric due to multiple valence band extrema close in energy. These qualitative ideas will be tested by explicit calculations of thermopower in the next section.

\subsection{Thermoelectric properties}
From the analysis of the DOS, the sharp increase in the DOS at the band edge suggests that the investigated compounds may have good thermoelectric properties, particularly large thermopower. To further explore this, we have studied the thermoelectric properties of both the marcasite and pyrite FeX$_2$ using Boltzmann transport equation as implemented in BOLTZTRAP code.\cite{Madsen} All the properties are calculated using Rigid Band Approximation (RBA)\cite{Scheidemantel,Jodin,Chaput} and the relaxation time $\tau$ is assumed to be independent of energy.\cite{singh, aggate2, Khuong} In Table III we see that the effective masses change with symmetry directions for both m and p structures. Since most of the experiments are done in poly-crystalline samples, we have calculated the average of thermopower and conductivity over three orthogonal axes. The calculated thermoelectric properties such as thermopower (S in $\mu$ V/K), electrical conductivity ($\frac{\sigma}{\tau}$ in $\Omega^{-1} m^{-1} s^{-1}$) and power factor scaled by $\tau$ ($\frac{S^2\sigma}{\tau}$ in $W/mK^2s$) for both the electron (n$_e$) and hole (n$_h$) dopings are given in Fig. 6-9. The melting temperatures of the marcasite phase of both the compounds are around 850 K, so we have calculated these properties up to 800 K for this structrure. The pyrite structure on the other hand is found to be stable up to 1300 K, and we have calculated S and $\frac{S^2\sigma}{\tau}$ up to 1200 K.

\par The observed reduction in the absolute value of the thermopower with decrease in the concentration is a peculiar feature of bipolar conduction (both electrons and holes contribute significantly to transport) at fixed temperature which we have seen in the case of p-FeSe$_2$ (Fig.7), m- and p-FeTe$_2$ (Fig.8 and 9) which is due to the small band gaps of these compounds (see Table II). From Fig. 6, we find that in m-FeSe$_2$ the thermopower values are almost similar for both electron and hole doping, whereas the electrical conductivity is higher in the case of hole doping in comparison with the electron doping. A similar behaviour is also seen in the power factor values. For p-FeSe$_2$, we have seen that up to $\sim$ 600 K all the thermoelectric quantities are better in the hole doping case which is also consistent with the results of DOS, but at high temperatures (800 K and 1000K) we find evidence of bipolar conduction. So p-FeSe$_2$ can be a good thermoelectric up to 600 K. In the case of m-FeTe$_2$ we find that electron doping is more favourable compared to the hole doping, whereas in p-FeTe$_2$ hole doping is favourable compared to electron doping. We find evidence of bipolar conduction in m-FeTe$_2$ above at 600 K and p-FeTe$_2$ above 800 K. So both m- and p-FeTe$_2$ can be used as thermoelectric material below 600 K.

\par As per the earlier study, the optimum value of the magnitude of $S$ usually falls in the region of 200-300 $\mu$ V/K to get a figure of merit (ZT) to be $\sim$1. In our study the hole concentration is between 2.10$\times 10^{19} - 7.96 \times 10^{19} cm^{-3}$, 1.78$\times 10^{20} - 5.56\times 10^{20} cm^{-3}$ for m- and p-FeSe$_2$ respectively. In case of FeTe$_2$ the optimum value in the marcasite phase is found in the electron concentration range  of 1.46$\times 10^{19} - 5.40\times 10^{20} cm^{-3}$ and for the pyrite phase it is found for the hole concentration range 1.36$\times 10^{20} - 5.31\times 10^{20} cm^{-3}$ at 600 K.

\par Our theoretical results for $S$ are compared with the earlier experimental work of Harada,\cite{Harada} and can be compared with the thermopower values at room and high temperature for the m-FeSe$_2$ structures. For marcasite the hole and electron concentrations are found to be 5.8$\times 10^{18} cm^{-3}$ and 8.5$\times 10^{19} cm^{-3}$ for a thermopower of +320 $\mu V/K$ and -120 $\mu V/K$ at 300 and 600 K, respectively. Similarly, for m-FeTe$_2$ we find the hole and electron concentration to be 9.2$\times 10^{19} cm^{-3}$ and 1.4$\times 10^{21} cm^{-3}$ for the thermopower values of 96 $\mu V/K$ and -74 $\mu V/K$ at 300 and 600 K. The experimental data on thermoelectric power and electrical conductivity is used to obtain an estimation of the relaxation time $\tau$.
We find $\tau$ to be 1.01$\times 10^{-15}$ {\it s} and 2.38$\times 10^{-15}$ {\it s} for FeSe$_2$ at 300 and 600 K respectively. Similarly, for m-FeTe$_2$ $\tau$ is found to be 2.3$\times 10^{-14}$ {\it s}  and 3.1$\times 10^{-14}$ {\it s} respectevely. We can clearly see that the relaxation time of m-FeSe$_2$ is lower that that of m-FeTe$_2$, and hence one could expect that FeSe$_2$ shows better TE properties than FeTe$_2$. Overall, both marcasite and pyrite phases of the investigated compounds are good candidates for thermoelectric properties, and marcasite FeSe$_2$ is found to be the best thermoelectric material among all the compounds studied. In order to evaluate these compound's figure of merit ZT, one should have experimental measurements of their thermal conductivities.
 \section{Conclusion}
The structural and electronic transport properties of marcasite and pyrite phases of FeX$_2$ are studied using density functional theory. We didn't find any structural transition between the marcasite and pyrite, and also found that marcasite structure of both the compounds are energetically stable than the pyrite structure. The calculated ground state properties of FeX$_2$(X=Se,Te) agree quite well with the available experiments. Electronic structure calculations show that all the investigated compounds are indirect band gap semiconductors, in good agreement with earlier reports. We further calculated the thermoelectric properties of the these compounds and compared with the available experimental reports. The calculations show all the investigated compounds to be very good thermoelectric materials for p-type doping, except marcasite FeTe$_2$ which favours electron doping. Among all the studied compounds we find marcasite FeSe$_2$ to be a good p-type thermoelectric material.

\section{Acknowledgements}
V. K. G and V. K would like to acknowledge IIT-Hyderabad for the computational facility. V. K. G. would like to thank MHRD for the fellowship. V. K. thank NSFC awarded Research Fellowship for International Young Scientists under Grant No. 11250110051. One of us (SDM) would like to acknowledge support by the Department of Energy (DOE)-Energy Frontier Research Center(EFRC) at Michigan State University on “Revolutionary Materials for Solid State Energy Conversion”.

\clearpage
*Author for Correspondence, E-mail: {\it kanchana@iith.ac.in}

\newpage

\begin{table}
\caption{Ground state properties of FeX$_2$(X=Se,Te) with GGA functional along with the available experimental results.}
\begin{tabular}{ccccccccccccccccccccccccccc}
\hline
&	&	&FeSe$_2$	&		&	&	&&&FeTe$_2$\\
\hline
		&Marcasite	&&&Pyrite	&	&	&&Marcasite	&	&Pyrite\\
\hline
&	Present	&Exp.$^a$ &&Present &Exp.$^b$		&&Present &Exp.$^c$ &&Present &Exp.$^b$	\\

a(\AA)	&4.7627		&4.8002	&&5.746		&5.7859	&&5.2845	&5.275	&&6.3083	&6.2937	\\

b(\AA)	&5.7439		&5.7823	&&		&	&&6.2865	&6.269	&&		&	\\	

c(\AA) 	&3.5872		&3.5834	&&		&	&&3.9058	&3.872	&&		&	\\

V(\AA$^3$)	&98.13	&99.46	&&189.71&193.70	&&129.75	&128.04	&&251.04	&249.30	\\


\hline
$a:$ Ref. \onlinecite{Arne}\\
$b:$ Ref. \onlinecite{TAB}\\
$c:$ Ref. \onlinecite{Yamaguchi}
\end{tabular}
\end{table}

\begin{table}
\caption{Band Gaps of marcasite and pyrite FeX$_2$(X=Se,Te) along with available experimental results in eV}
\begin{tabular}{cccccccccccccc}
\hline
	&	&FeSe$_2$	&		&	&	&&&FeTe$_2$\\
\hline
		&Marcasite	&&Pyrite	&	&	&&Marcasite	&	&Pyrite\\
\hline 	
Present         &1.234		&&0.694		&	&	&&0.328		&	&0.432	\\

Exp/other 	&0.95-1.03$^a$	&&-		&	&	&&0.92$^c$	&	&-	  \\

Other calculation&0.86$^b$	&&0.67$^b$	&	&	&&-		&	&-	\\
\hline
$a:$ Ref. \onlinecite{Arne}\\
$b:$ Ref. \onlinecite{Ganga}\\
$c:$ Ref. \onlinecite{Landolt}
\end{tabular}
\end{table}

\clearpage
\newpage
\begin{table}
\caption{The calculated effective mass of the marcasite and pyrite of both FeSe$_2$ and FeTe$_2$ in some selected directions of the Brillouin zone in the units of electron rest mass.}
\begin{tabular}{cccccccccccccc}
\hline
Marcasite		&FeSe$_2$		&	&&FeTe$_2$ \\
\hline
Direction  &Valence Band	&Conduction Band	&&Valence Band	&Conduction Band \\
$\Gamma$-Z	&0.048		&0.451			&&0.019		&0.038	\\
$\Gamma$-Y	&0.042		&0.041			&&0.018		&0.021	\\
$\Gamma$-X	&0.024		&0.066			&&0.017		&0.014	\\
\hline
Pyrite \\
$\Gamma$-X	&0.012		&0.028			&&0.010		&0.027	\\
$\Gamma$-M	&0.028		&0.055			&&0.019		&0.046	\\
$\Gamma$-R	&0.032		&0.036			&&0.028		&0.025	\\
\hline
\end{tabular}
\end{table}
\clearpage
\newpage

\begin{figure}
\begin{center}
\subfigure[]{\includegraphics[width=60mm,height=60mm]{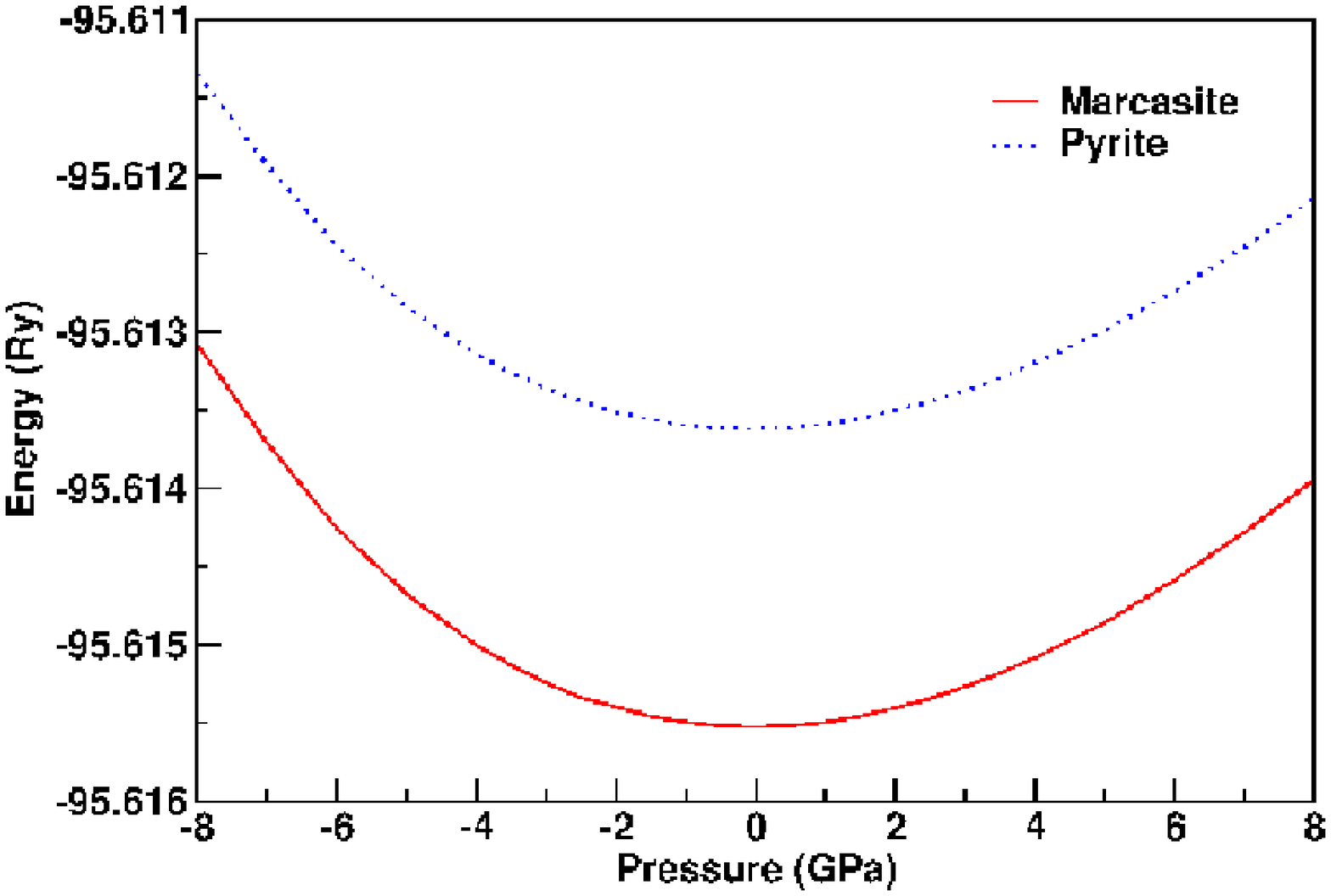}}
\subfigure[]{\includegraphics[width=60mm,height=60mm]{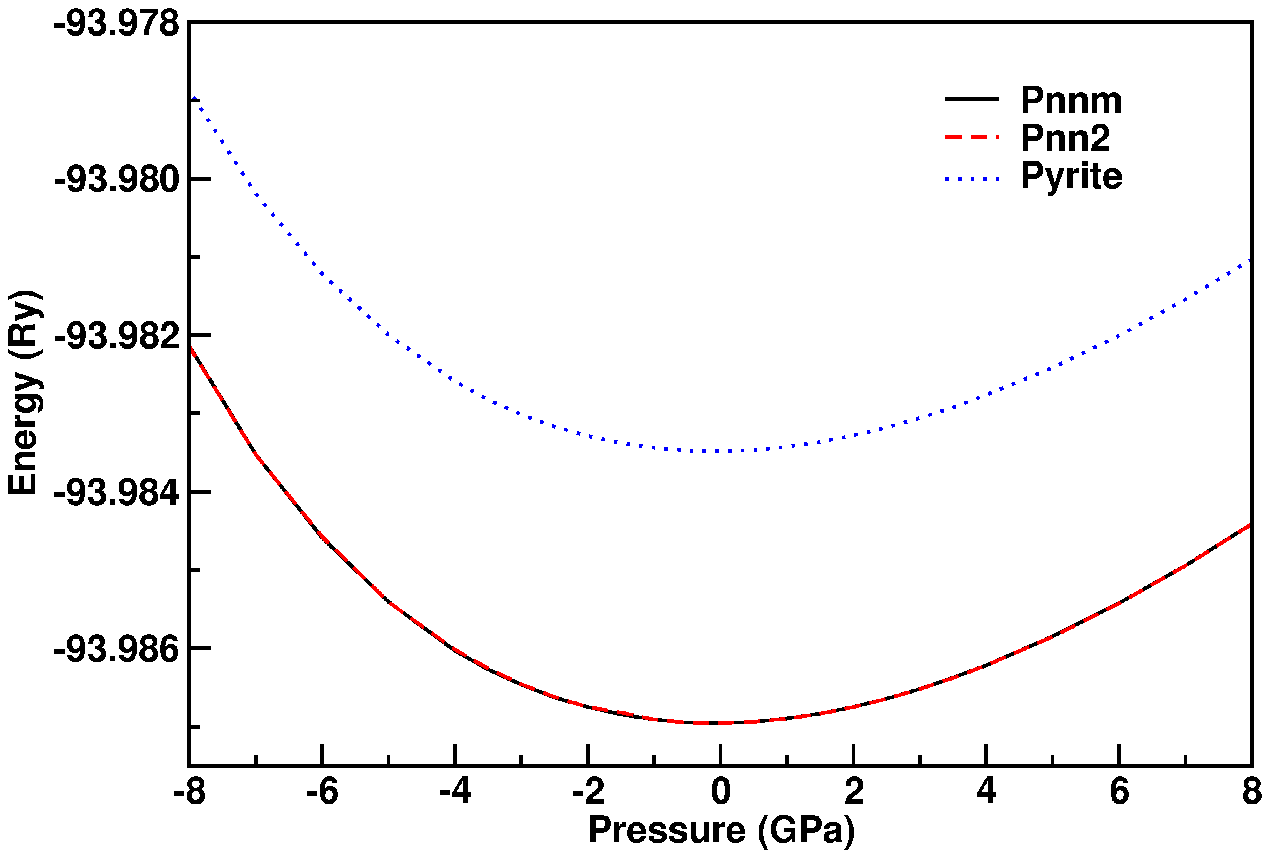}}
\caption{(Color online) Variation of the total energy with pressure (a) FeSe$_2$ (b) FeTe$_2$}
\end{center}
\end{figure}

\begin{figure}
\begin{center}
\subfigure[]{\includegraphics[width=80mm,height=80mm]{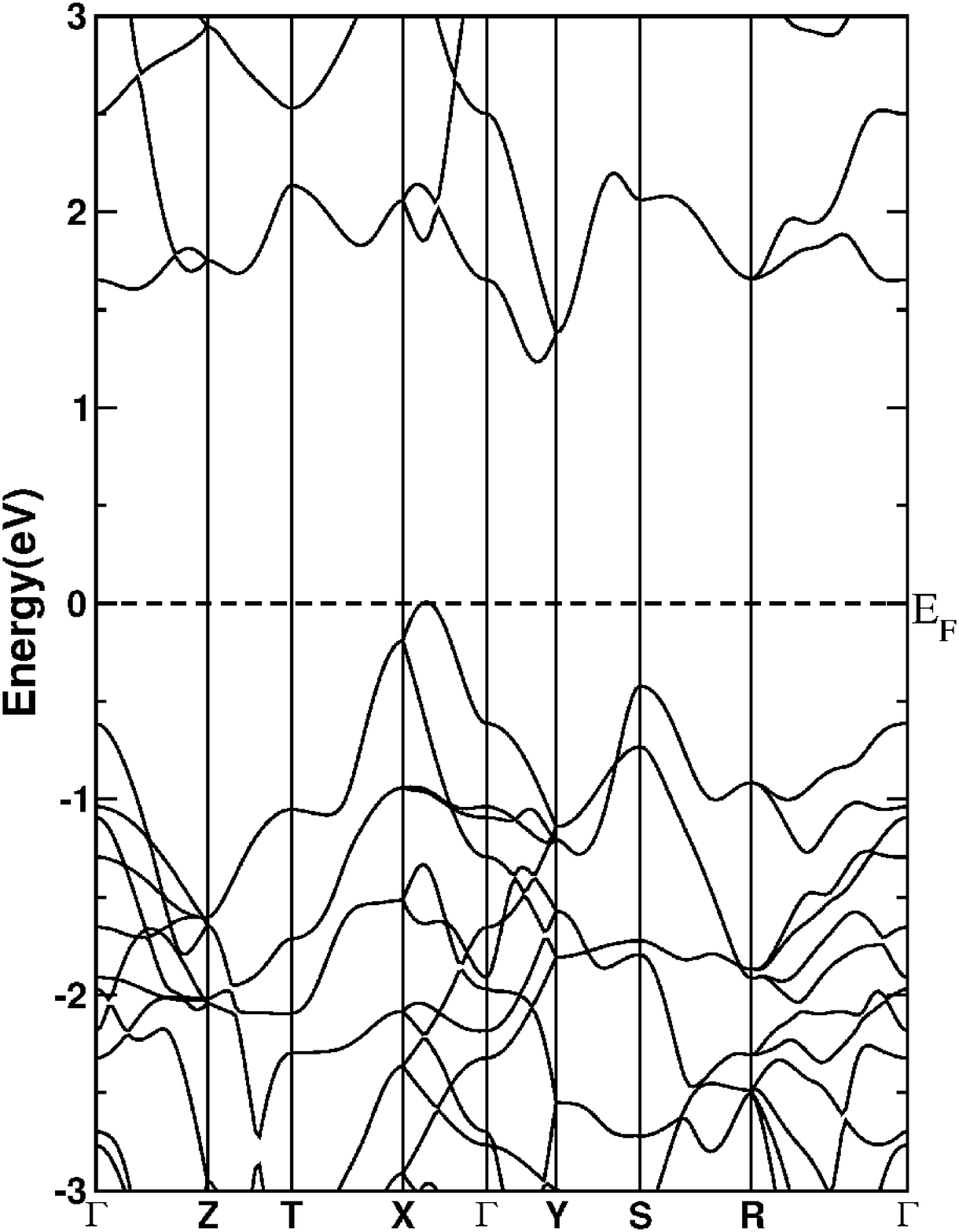}}
\subfigure[]{\includegraphics[width=80mm,height=80mm]{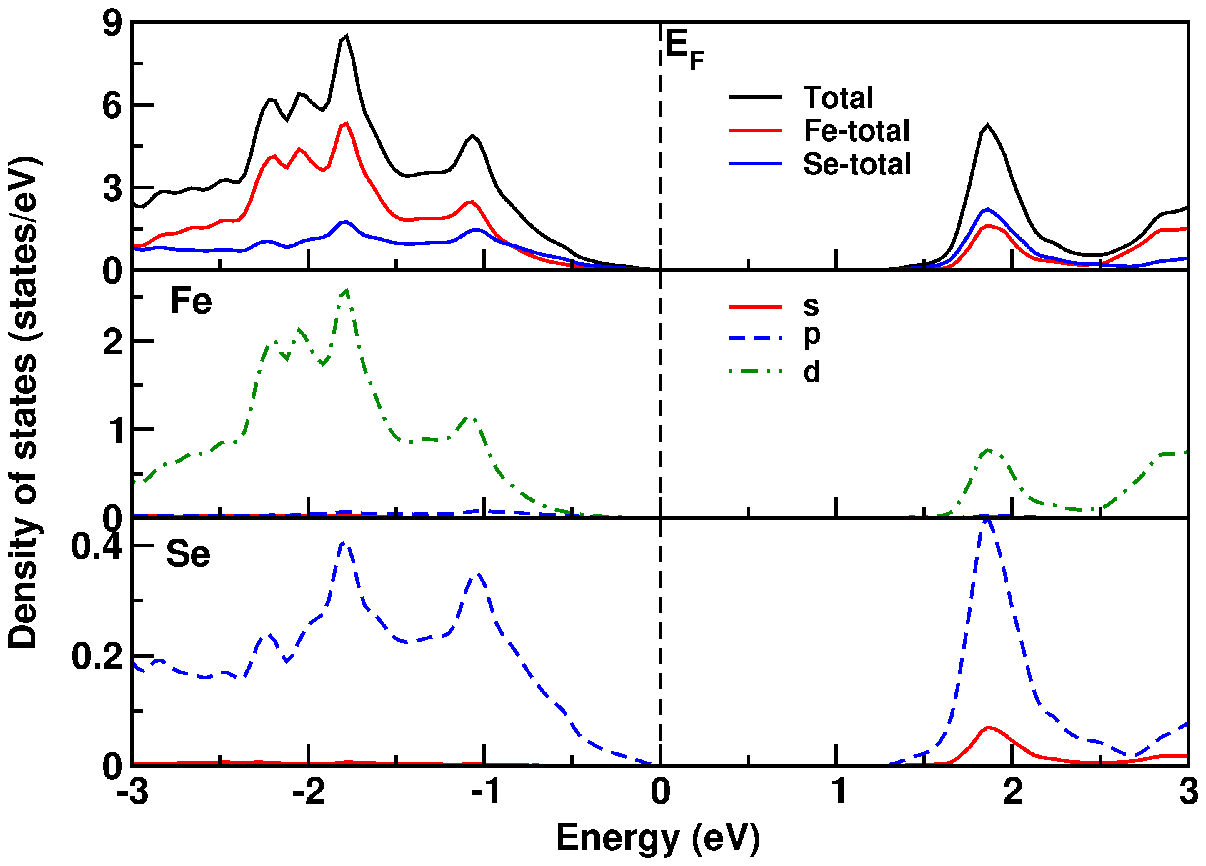}}
\caption{(Color online) (a) Band structure (b) Density of states of marcasite FeSe$_2$ within the exchange correlation of GGA+U with a value of U$_{Fe}$= 0.52 Ry as implemented in WIEN2k\cite{Liechtenstein} code at theoretical equilibrium volume.}
\end{center}
\end{figure}

\begin{figure}
\begin{center}
\subfigure[]{\includegraphics[width=80mm,height=80mm]{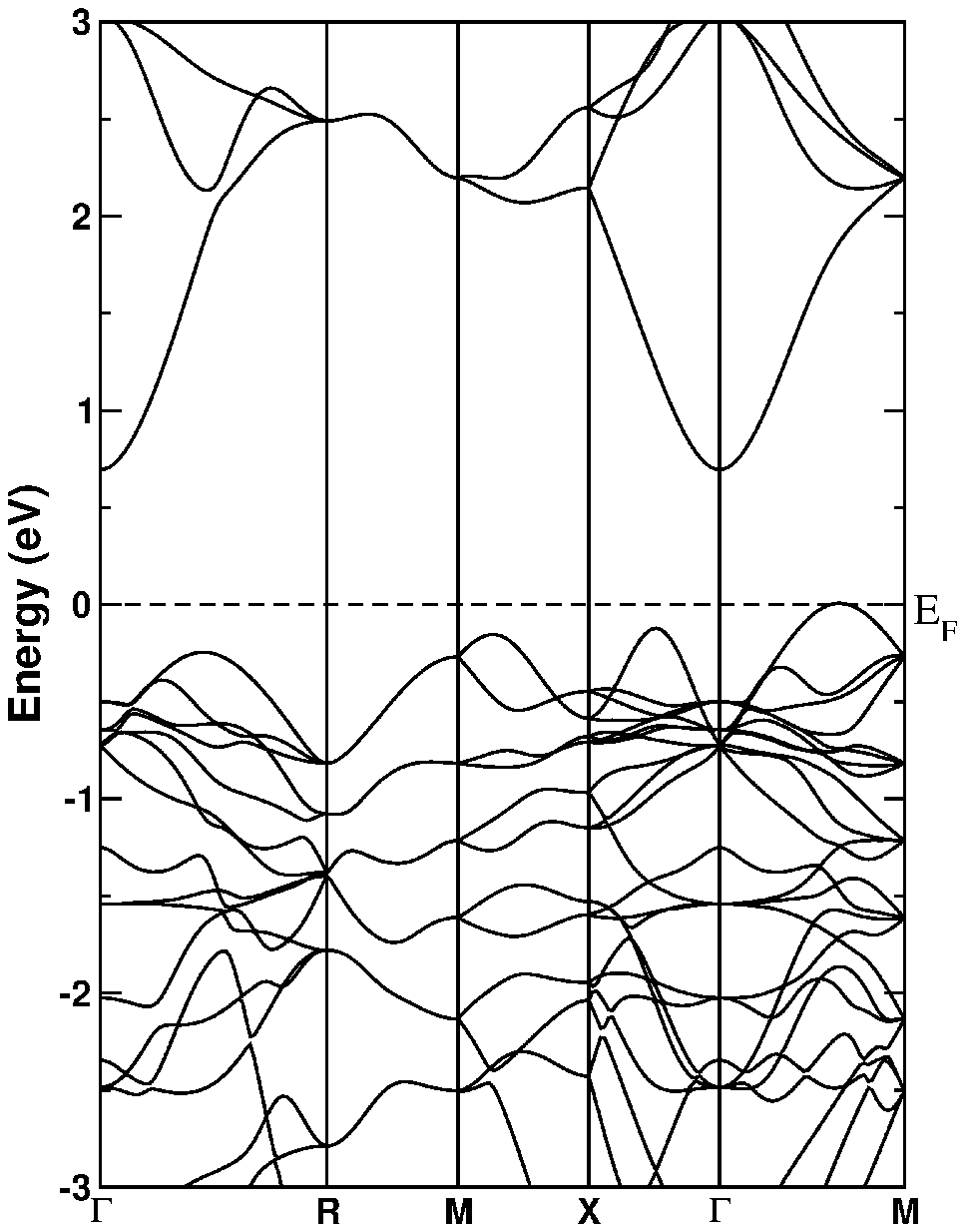}}
\subfigure[]{\includegraphics[width=80mm,height=80mm]{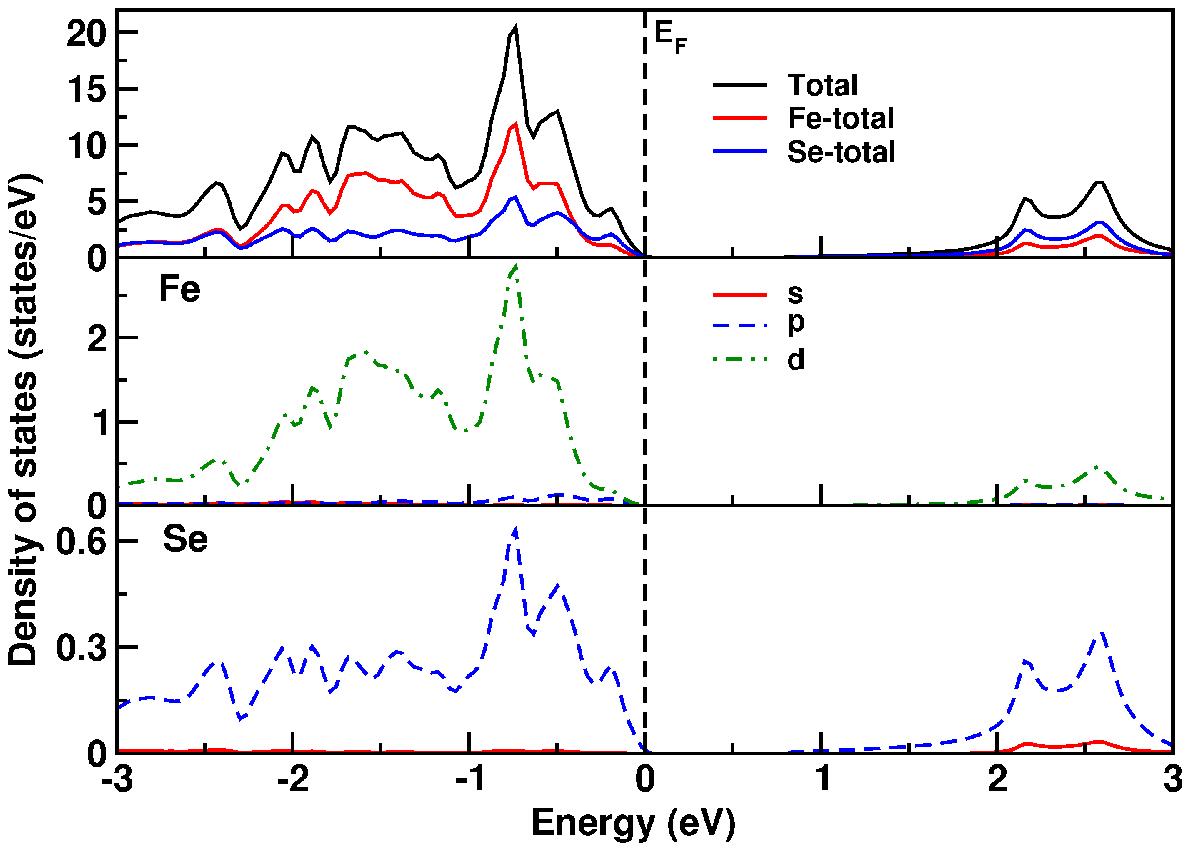}}
\caption{(Color online) (a) Band structure (b) Density of states of pyrite FeSe$_2$ within the exchange correlation of GGA+U with a value of U$_{Fe}$= 0.52 Ry as implemented in WIEN2k\cite{Liechtenstein} code at theoretical equilibrium volume.}
\end{center}
\end{figure}

\clearpage
\newpage

\begin{figure}
\begin{center}
\subfigure[]{\includegraphics[width=80mm,height=80mm]{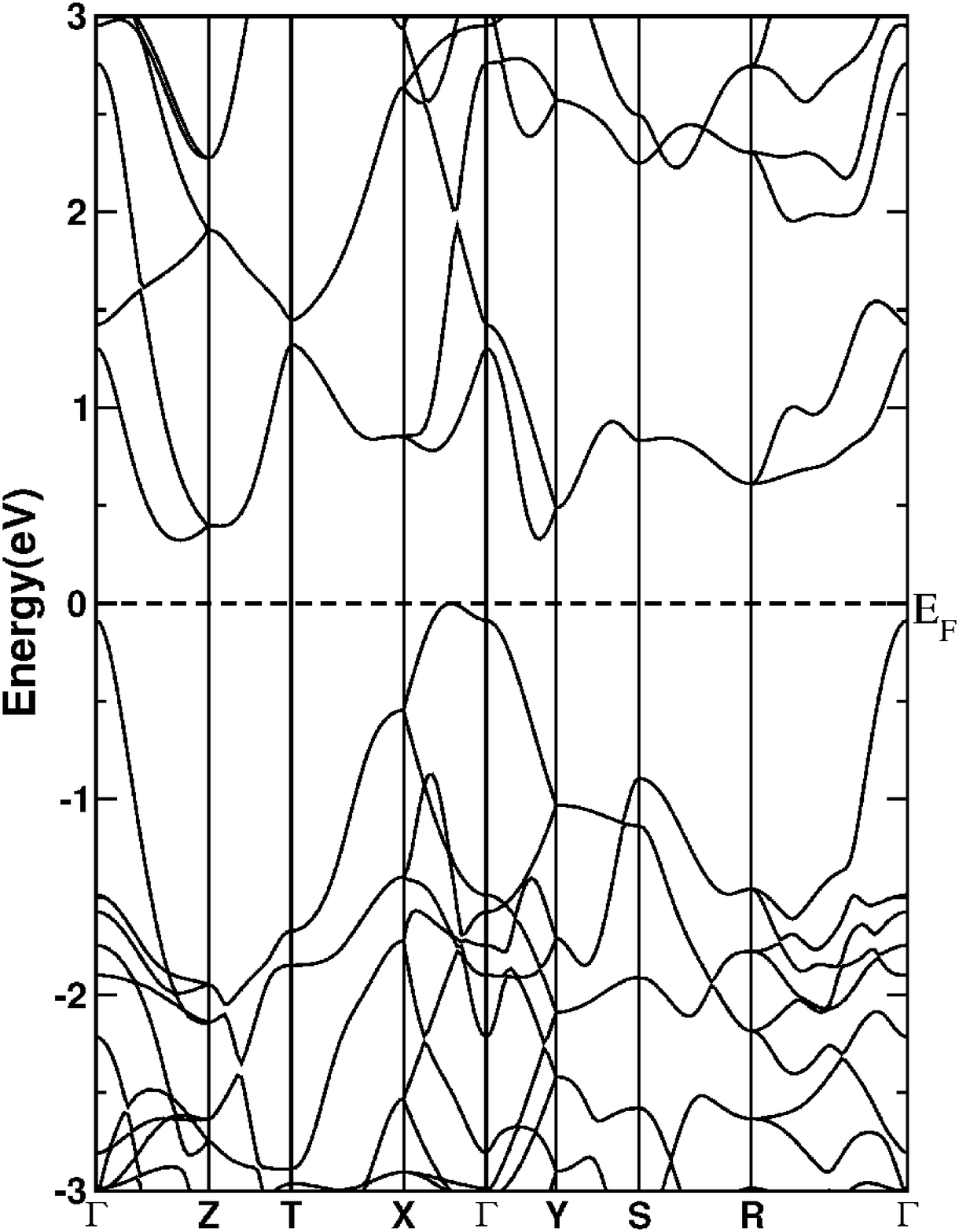}}
\subfigure[]{\includegraphics[width=80mm,height=80mm]{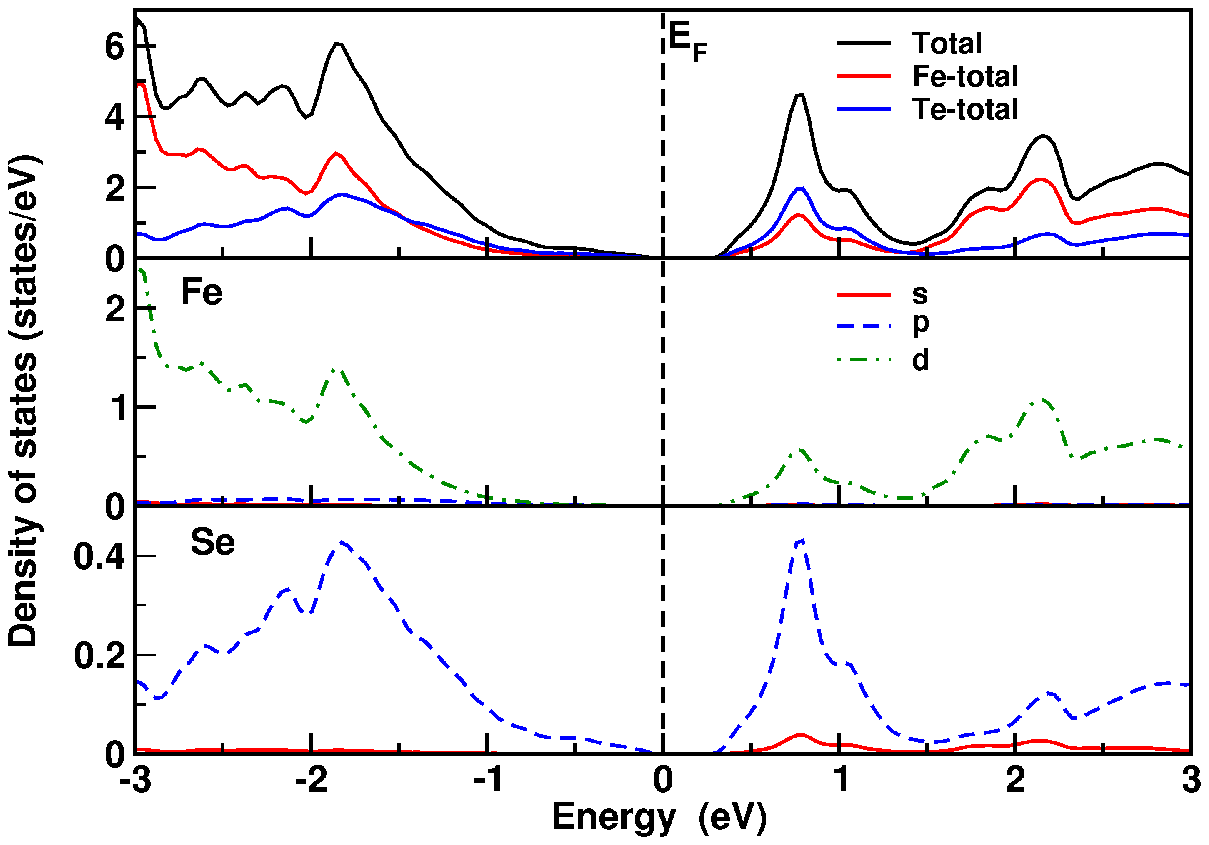}}
\caption{(Color online) (a) Band structure (b) Density of states of marcasite FeTe$_2$ within the exchange correlation of GGA+U with a value of U$_{Fe}$= 0.52 Ry as implemented in WIEN2k\cite{Liechtenstein} code at theoretical equilibrium volume.}
\end{center}
\end{figure}

\begin{figure}
\begin{center}
\subfigure[]{\includegraphics[width=80mm,height=80mm]{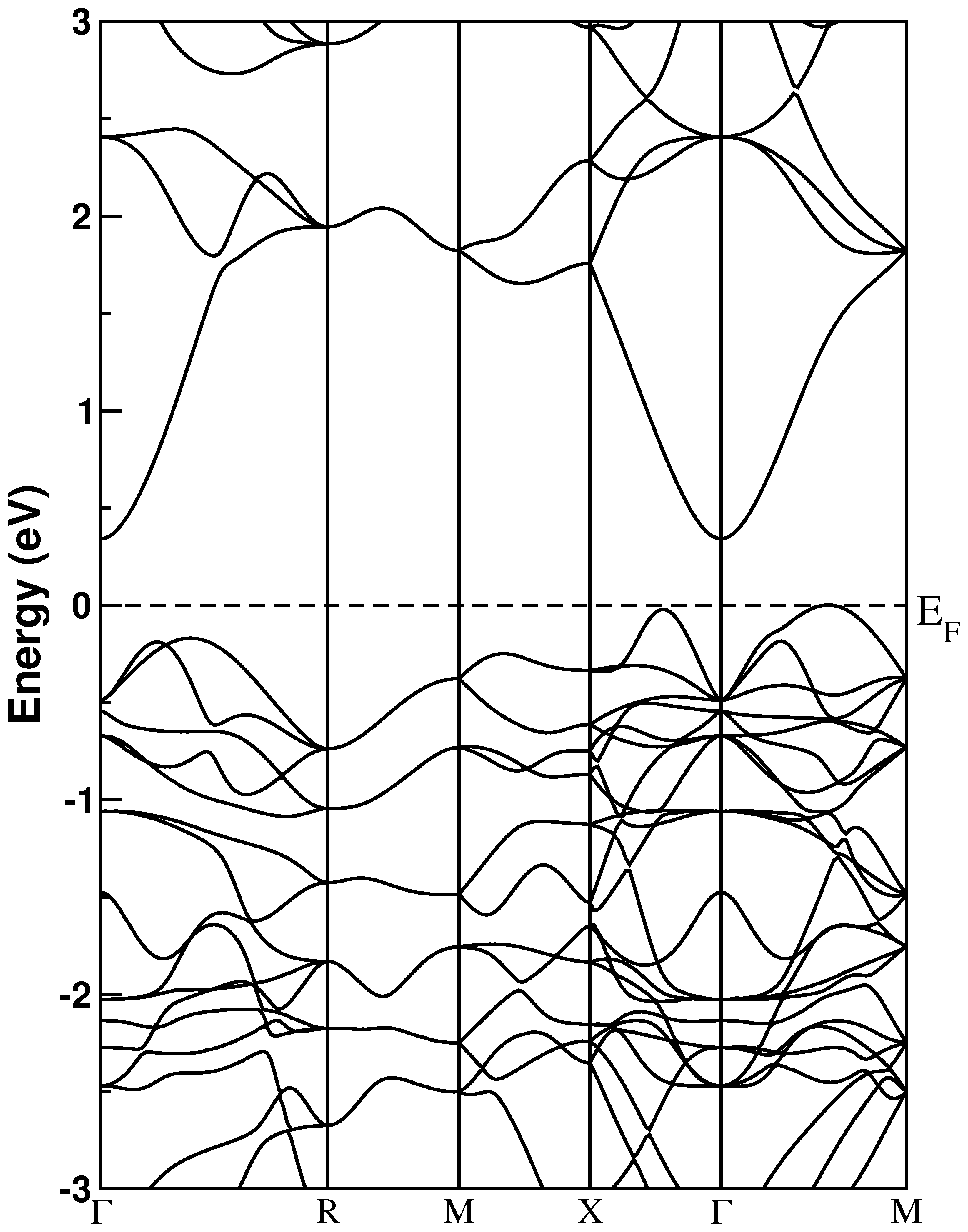}}
\subfigure[]{\includegraphics[width=80mm,height=80mm]{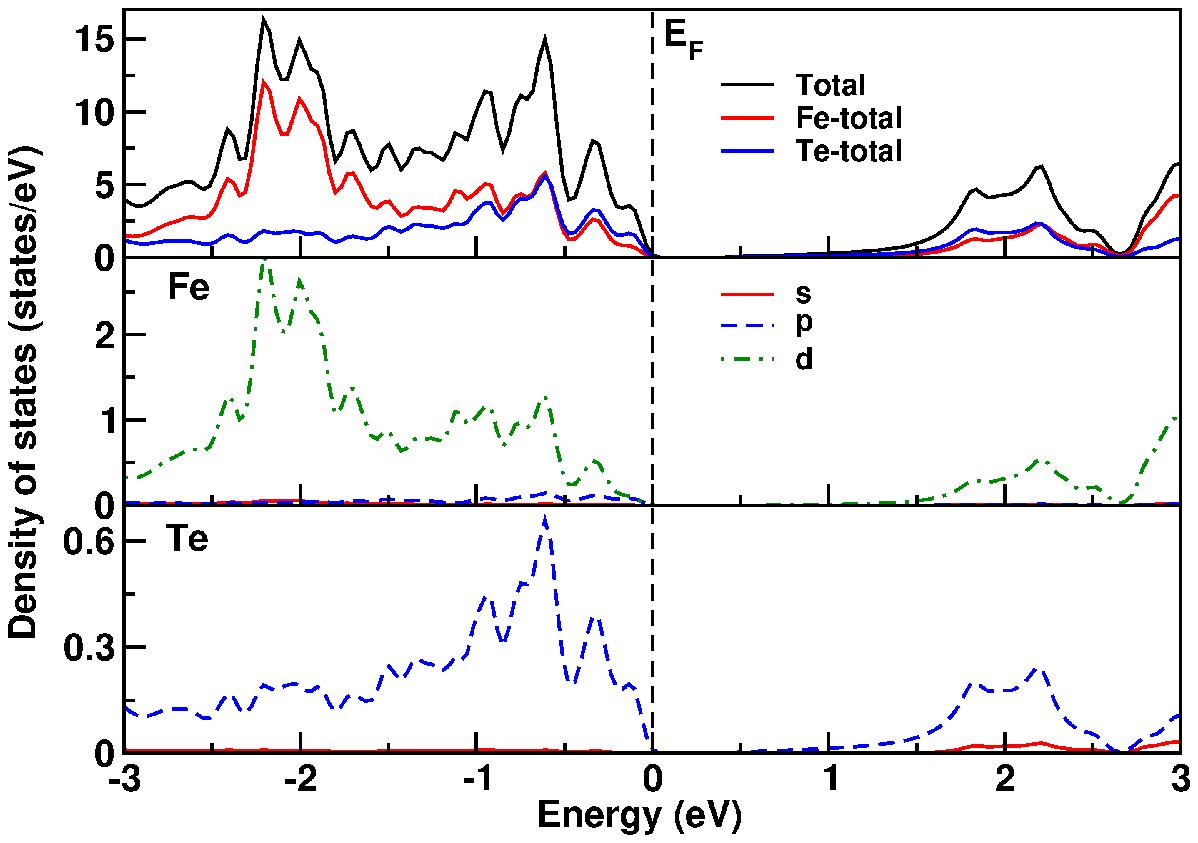}}
\caption{(Color online) (a) Band structure (b) Density of states of pyrite FeTe$_2$ within the exchange correlation of GGA+U with a value of U$_{Fe}$= 0.52 Ry as implemented in WIEN2k\cite{Liechtenstein} code at theoretical equilibrium volume.}
\end{center}
\end{figure}

\clearpage
\newpage

\begin{figure}
\begin{center}
\includegraphics[width=90mm,height=90mm]{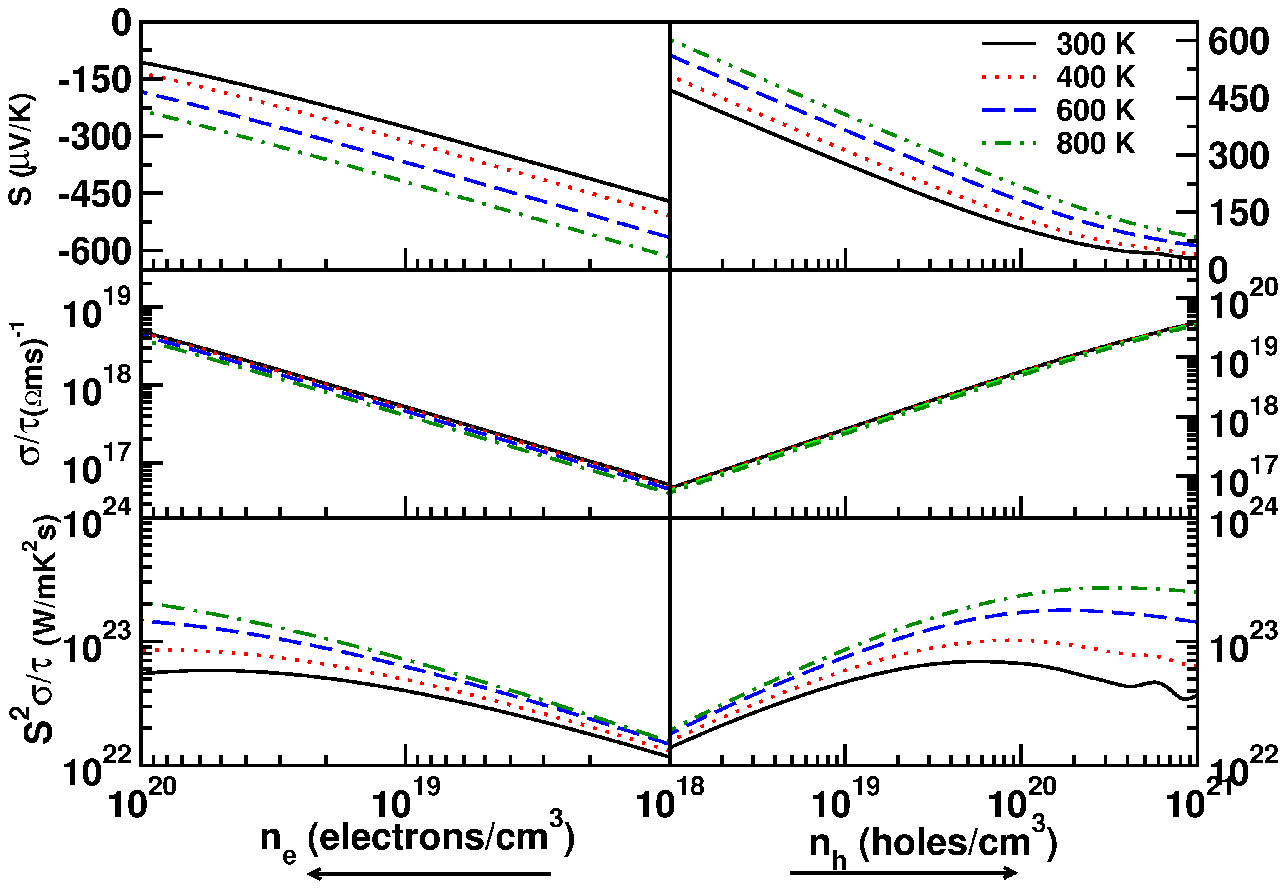}
\caption{(Color online) Thermoelectric properties of thermopower($S$), electrical conductivity scaled by relaxation time($\sigma/\tau$) and power factor scaled by relaxation time($S^2 \sigma/\tau$) for both electron(left) and hole(right) doping of marcasite FeSe$_2$}
\end{center}
\end{figure}

\clearpage
\newpage

\begin{figure}
\begin{center}
\includegraphics[width=90mm,height=90mm]{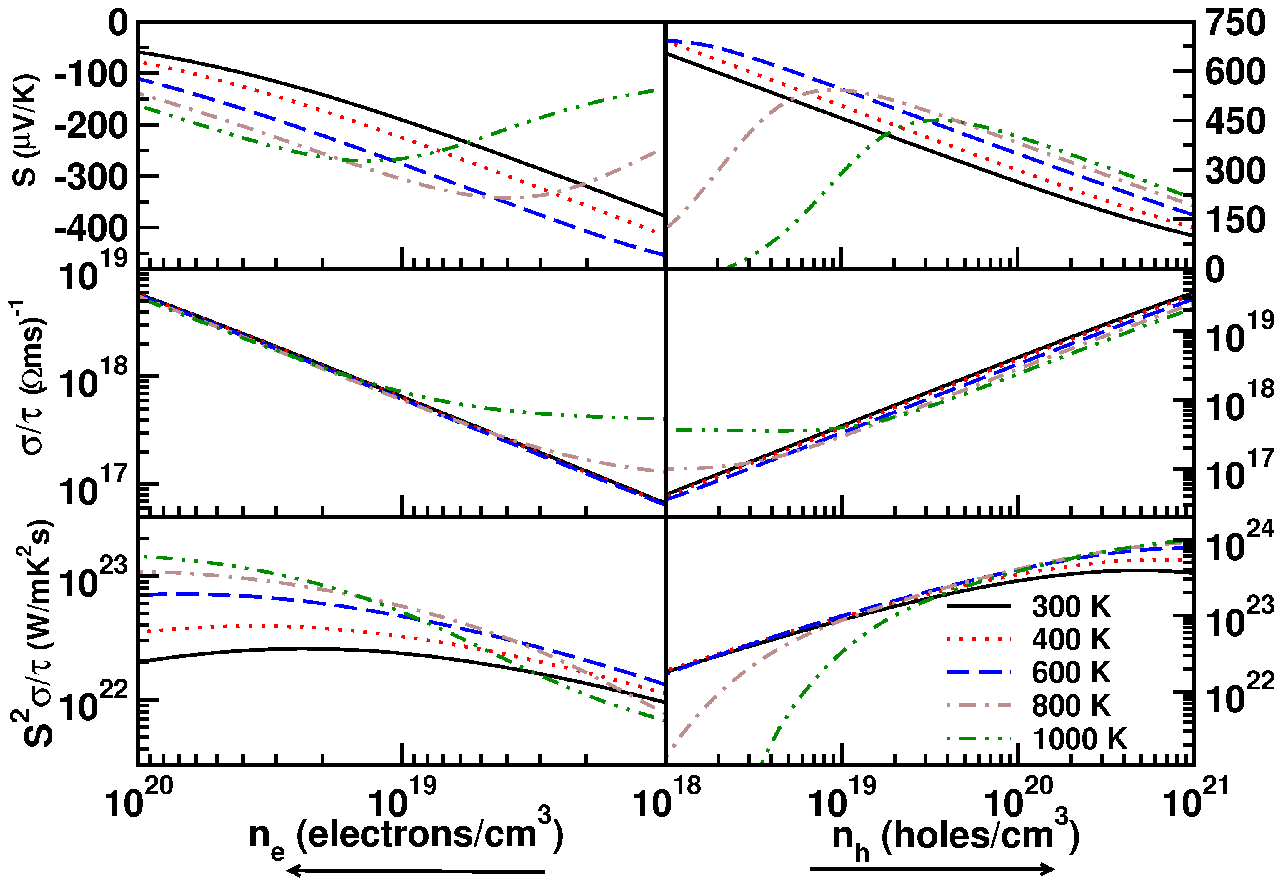}
\caption{(Color online) Thermoelectric properties of thermopower($S$), electrical conductivity scaled by relaxation time($\sigma/\tau$) and power factor scaled by relaxation time($S^2 \sigma/\tau$) for both electron(left) and hole(right) doping of pyrite FeSe$_2$}
\end{center}
\end{figure}

\clearpage
\newpage

\begin{figure}
\begin{center}
\includegraphics[width=90mm,height=90mm]{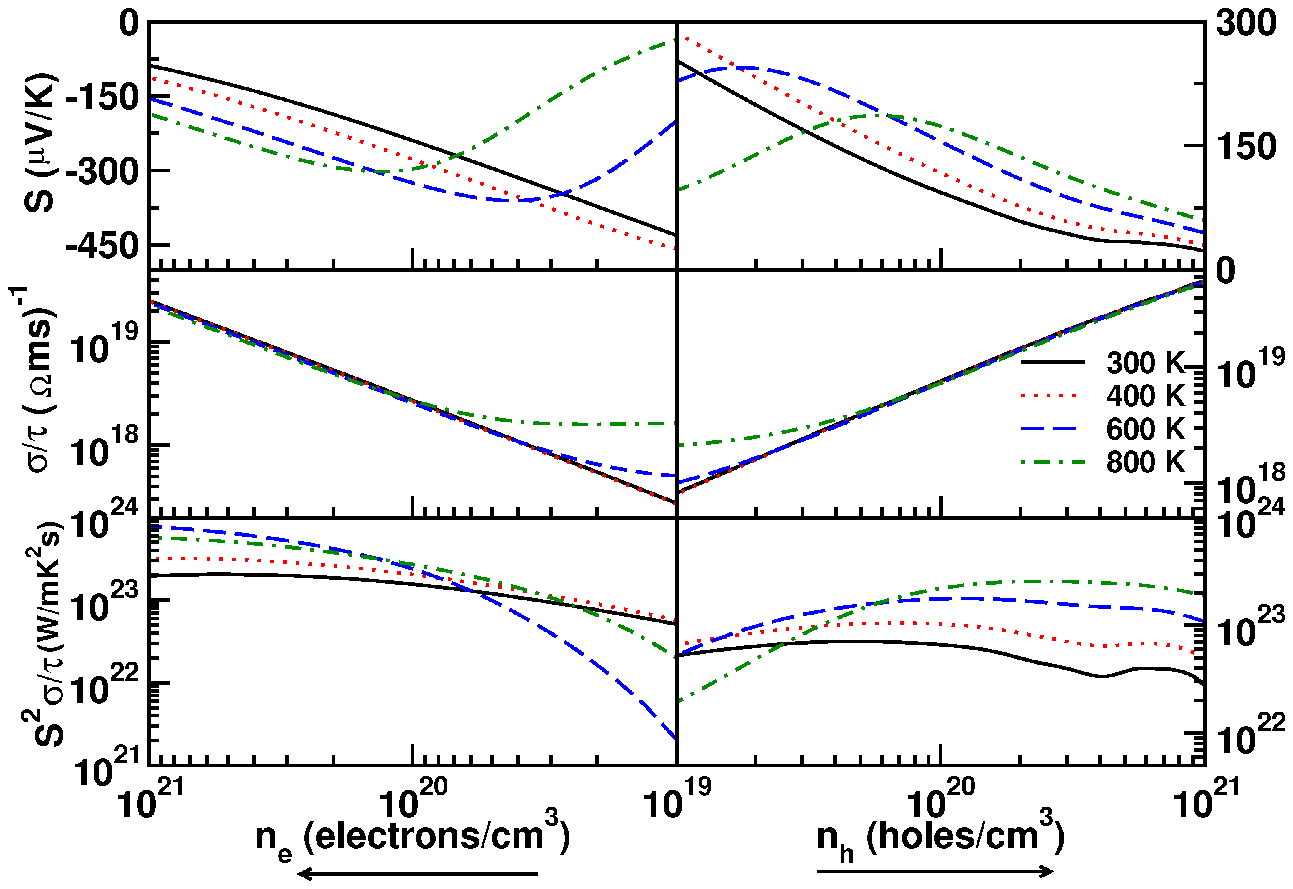}
\caption{(Color online) Thermoelectric properties of thermopower($S$), electrical conductivity scaled by relaxation time($\sigma/\tau$) and power factor scaled by relaxation time($S^2 \sigma/\tau$) for both electron(left) and hole(right) doping of marcasite FeTe$_2$}
\end{center}
\end{figure}

\clearpage
\newpage

\begin{figure}
\begin{center}
\includegraphics[width=90mm,height=90mm]{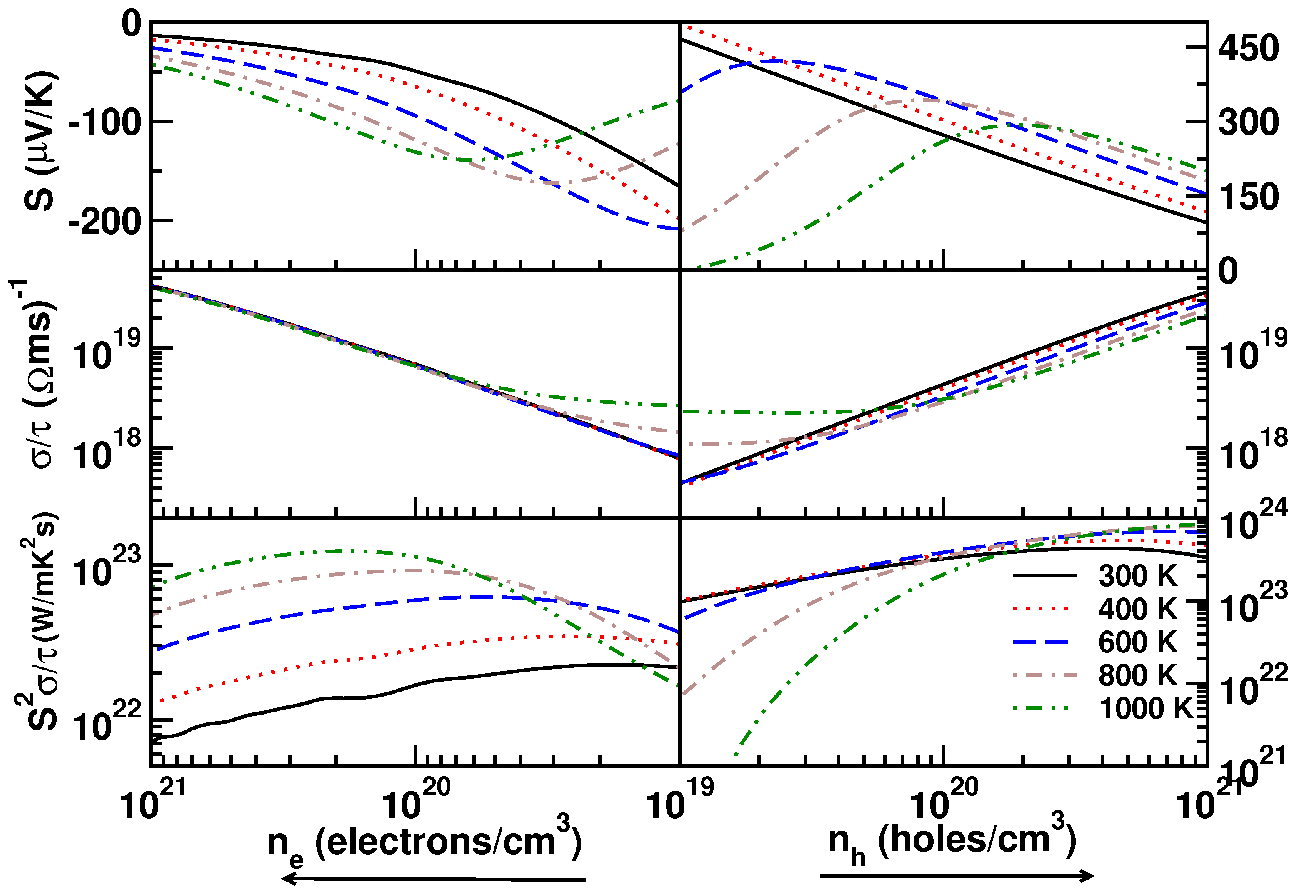}
\caption{(Color online) Thermoelectric properties of thermopower($S$), electrical conductivity scaled by relaxation time($\sigma/\tau$) and power factor scaled by relaxation time($S^2 \sigma/\tau$) for both electron(left) and hole(right) doping of pyrite FeTe$_2$}
\end{center}
\end{figure}

\end{document}